\documentclass[manuscript]{aastex}

\usepackage{tikz}
\usepackage{amsmath}

\usepackage{graphicx}

\usepackage{pdflscape}

\slugcomment{\textbf{The Astronomical Journal, in press}}

\shorttitle{Active Asteroid 6478 Gault}
\shortauthors{Jewitt}

\begin{document}

\title{Episodically Active Asteroid 6478 Gault}

%

\author{David Jewitt$^{1,2}$, Yoonyoung Kim$^3$, Jane Luu$^{4}$,  Jayadev Rajagopal$^5$, Ralf Kotulla$^6$, Susan Ridgway$^5$, Wilson Liu$^5$
}

\affil{$^1$ Department of Earth, Planetary and Space Sciences,
UCLA, 
595 Charles Young Drive East, 
Los Angeles, CA 90095-1567\\
$^2$ Dept.~of Physics and Astronomy,
UCLA, 
430 Portola Plaza, Box 951547,
Los Angeles, CA 90095-1547\\
$^3$  Max Planck Institute for Solar System Research, Justus-von-Liebig-Weg 3, 37077 G\"ottingen, Germany\\
$^4$ Department of Physics and Technology, Arctic University of Tromso, Tromso, Norway \\
$^5$ NOAO, 950 North Cherry Ave., Tucson, AZ 85719 \\
$^6$ Department of Astronomy, University of Wisconsin-Madison, 475 N. Charter St., Madison, WI 53706 \\
}

\email{jewitt@ucla.edu}

\begin{abstract}
We present imaging and spectroscopic observations of  6478 Gault, a $\sim$6 km diameter inner  main-belt asteroid currently exhibiting strong, comet-like characteristics.  Three distinct tails indicate that ultra-slow dust (ejection speed 0.15$\pm$0.05 m s$^{-1}$) was emitted from Gault in separate episodes beginning UT 2018 October 28$\pm$5 (Tail A),  UT 2018 December 31$\pm$5 (Tail B), and UT 2019 February 10$\pm$7 (Tail C), with durations of $\Delta T \sim$ 10 to 20 days.   With a mean particle radius $\overline{a} \sim$ 200 $\mu$m, the estimated masses of the  tails are $M_A \sim 4\times10^{7}$ kg, $M_B \sim 6\times 10^{6}$ kg and $M_C \sim 6\times10^5$ kg, respectively, and the mass loss rates from the nucleus are 20 to 40 kg s$^{-1}$ for Tail A, 4 to 6 kg s$^{-1}$ for Tail B and $\sim$0.4 kg s$^{-1}$ for Tail C.  In its optical colors Gault is more similar to C-type asteroids than to S-types, even though the latter are numerically dominant in the inner asteroid belt.  A spectroscopic upper limit to the production of gas is set at 1 kg s$^{-1}$.  Discrete emission in three protracted episodes effectively rules out an impact origin for the observed activity.  Sublimation driven activity is unlikely given the inner belt orbit and the absence of detectable gas. In any case, sublimation would not easily account for the observed multiple ejections.  The closest similarity is between Gault and active asteroid 311P/(2013 P5), an object showing repeated but aperiodic ejections of dust over a 9 month period.  While Gault is 10 times larger than 311P/(2013 P5), and the relevant timescale for spin-up by radiation torques is $\sim$100 times longer, its properties are likewise most consistent with episodic emission from a body rotating near breakup.  
\end{abstract}


\section{INTRODUCTION}

Main-belt asteroid 6478 Gault (formerly 1988 JC1 and simply ``Gault'', hereafter) was first reported to be active in observations from the ATLAS telescope on UT 2019 January 8,  when it sported a long, thin tail (Smith et al.~2019).  Soon, archival observations from the ZTF survey emerged to show dust emission from Gault starting as early as October and with peaks in the brightness near October 31 and December 30 (Ye et  al.~2019).  The orbital elements (semimajor axis $a$ = 2.305 AU, eccentricity $e$ = 0.194 and inclination $i$ = 22.8\degr), qualify Gault as a member of the Phocaea asteroid group, the latter being bounded by 2.1 $< a < $2.5 AU, 0.0 $< e < 0.4$ and 17\degr $< i <$ 30\degr~(Novakovi{\'c} et al.~2017).  If substantiated, this association may suggest that Gault is a fragment from a 260$\pm$40 Myr old shattering collision.  The inner belt orbit of Gault is near to  active asteroids 354P/(2010 A2) and 311P/(2013 P5).  Of these, the former is a product of hypervelocity impact onto a $\sim$100 m scale primary (Jewitt et al.~2010, 2013a, Kim et al.~2017a, 2017b).  The cause of episodic mass loss in the latter is less clear, but  likely reflects incipient rotational instability (Jewitt et al.~2013b, 2015, Hainaut et al.~2014, Hirabayashi et al.~2015) and 311P may also be an eclipsing binary (Jewitt et al.~2018).  Several active asteroids in the outer-belt have been associated with repeated episodes of  near-perihelion mass loss, leading to the suggestion that they are driven by the sublimation of near surface water ice (Hsieh and Jewitt 2006).  The diversity of origins of the activity in otherwise dynamically unremarkable asteroids leads to the inevitable question ``what drives the activity in Gault?".  

In this paper, we present imaging and spectroscopic observations taken to characterize the on-going mass loss from this formerly unremarkable asteroid.

\section{OBSERVATIONS}
\subsection{Imaging}

We used the 2.56 m diameter Nordic Optical Telescope (NOT), located on La Palma, the Canary Islands and the Andalucia Faint Object Spectrograph and Camera (ALFOSC) optical camera.  ALFOSC is equipped with a 2048$\times$2064 pixel ``e2v Technologies'' charge-coupled device (CCD), giving an image scale of 0.214\arcsec~pixel$^{-1}$, and a vignette-limited field of view approximately 6.5\arcmin$\times$6.5\arcmin. Broadband Bessel B (central wavelength $\lambda_c$ = 4400\AA, full-width at half maximum (FWHM) $\Delta \lambda$ = 1000\AA), V ($\lambda_c$ = 5300\AA, $\Delta \lambda$ = 800\AA), and R ($\lambda_c$ = 6500\AA, $\Delta \lambda$ = 1300\AA) filters were used. The NOT was tracked at non-sidereal rates to follow Gault, while autoguiding using field stars. We obtained a series of images each of 70 s integration during which time field objects trailed by $\sim$0.01\arcsec~on January 10 but by $\sim$0.9\arcsec~on February 21 and $\sim$0.6\arcsec~on April 10. The images were first bias subtracted and then normalized by a flat-field image constructed from images of the illuminated interior of the observatory dome.  Photometric calibration of the data was obtained by reference to standard star PG 0918 + 029A (Landolt 1992). Measurements of field stars show that the atmosphere was photometric to $\lesssim$0.01 mag. The field stars were also used to measure the extinction, found to be 0.18, 0.12 and 0.06 magnitudes per airmass in the B, V and R filters, respectively.  Seeing was variable in the range 1.0\arcsec~FWHM to 1.3\arcsec~FWHM.   NOT data from February 21 suffered from strong internal scattering due to moonlight but this was largely removed in the processing of the data.

Additional observations were taken on UT 2019 March 4 at the 3.5 m diameter WIYN  telescope, located at Kitt Peak National Observatory in Arizona (see Table 1). We used the One Degree Imager camera mounted at the Nasmyth focus with a rebinned image scale of 0.25 pixel$^{-1}$ (Harbeck et al. 2014). Observations were taken through the Sloan r' (central wavelength $\lambda_c$= 6250\AA, FWHM = 1400\AA) filter with an accumulated integration of 6300 s (180 s individual integration time). Median seeing was 1.1\arcsec~FWHM. For data reduction we used the ``Quickreduce'' pipeline (Kotulla 2014), and photometric calibration of the data was made with reference to the Sloan DR 14 database (Blanton et al. 2017).

We also used publicly accessible observations from the Hubble Space Telescope (GO 15678, PI Meech) taken on UT 2019 February 05.  These consist of five images per visit, with 380 s exposures acquired through the F350 LP broadband filter (central wavelength 6230\AA, FWHM = 4758\AA) using the WFC3 camera.  The pixel size was 0.04\arcsec~pixel$^{-1}$.  The images were shifted to a common center and median-combined to eliminate cosmic rays to produce a composite image that was used for all measurements. Unfortunately, the nucleus of Gault is saturated in the HST data, which therefore cannot be used to study the near-nucleus environment at high resolution.  Here, we use the HST composites only to measure the tail position angles, as recorded in Table (\ref{geometry}).

\subsection{Photometry and Lightcurve}
\label{phot}

A composite image from UT 2019 January 10, formed by shifting the R-band images of combined exposure time 6790 s  to a common center, is shown in Figure (\ref{image_jan10}).  The asteroid shows a single, prominent tail (``Tail A'') at position angle PA = 291.0$\pm$0.5\degr~that is aligned with neither the anti-solar direction nor the projected negative heliocentric velocity vector.  A single image taken in the process of recording a spectrum on UT 2019 January 20  (see \textsection{\ref{spectra}}) showed this same tail (PA = 292$\pm$1\degr) plus a second, stubby tail at PA = 301$\pm$1\degr.  This second, stubby tail (``Tail B'')  is better recorded in NOT images from 
 UT February 21 (Figure \ref{image_feb21}) and WIYN on UT 2019 March 03 (Figure \ref{image_mar03}).  
These later images  show the second tail  shorter and brighter than the first, resulting from more recent ejection of dust.   By UT 2019 March 24 (Figure \ref{image_mar24}), when the Earth passed through the projected orbital plane of Gault, Tails A and B were merged.  Observations on UT 2019 April 10 from $\delta_E$ = -4\degr~below the plane show Tails A and B widely splayed, and the emergence of new Tail C (Figure \ref{image_apr10}).

We searched for evidence of dust between Tails A and B by making plots of the surface brightness perpendicular to the mid-line between the tails, using the high signal-to-noise ratio image from UT 2019 Mar 03 (Figure \ref{image_mar03}).  The search is compromised by  ultra-faint field galaxies whose trails are aligned roughly parallel to the tail mid-line, and no Gault-associated dust could be detected.  

Time-series measurements of the apparent magnitude of the near-nucleus region, $m_V$, were obtained from UT 2019 January 10, the data with the longest timebase. We used  circular apertures with angular radii 10, 15 and 20 pixels (2.1\arcsec, 3.2\arcsec~and 4.3\arcsec, respectively), with sky subtraction from a contiguous annulus having outer radius 8.4\arcsec.  Calibration against field and Landolt stars was made using the same apertures.   We used measurements of nearby field stars, obtained with the same photometric apertures, to assess variable atmospheric extinction as the airmass varied in the range 1.3 to 1.8 over the 3 hours of observation.  

In Figure (\ref{lightcurve}) we show a sample photometric time-series measured at  $\sim$2 minute cadence on UT 2019 January 10.  In a phase dispersion minimization analysis of the data, we find no evidence for cyclic (rotational) brightness variation at the $\sim$1\% level or larger.  The same invariance of the photometry  is observed using the 2.1\arcsec~and 3.1\arcsec~apertures.  This is in agreement with the findings of Moreno et al~(2019) and Ye et al.~(2019).  Kleyna et al.~(2019) report evidence for a $\sim$1 hour periodogram peak in longer time-series from several telescopes.  However, they find no convincing lightcurve, perhaps because of the influence of systematic uncertainties in the data, and/or weak and transient activity.  The lack of a measurable lightcurve is consistent with Gault having a shape that is close to azimuthally symmetric, or a rotation vector parallel to the line of sight, or with the scattering cross-section being dominated by dust.

To address this latter possibility, we computed the absolute magnitude from our data to compare with the absolute magnitude recorded in the JPL Horizons on-line database. The apparent magnitudes, $m_V$, and absolute magnitudes, $H$, are related by the inverse square law

\begin{equation}
H = m_V - 2.5\log_{10}\left(r_H^2 \Delta^2\right) +  2.5\log_{10}(\Phi(\alpha))
\label{abs}
\end{equation}

\noindent where $r_H$ and $\Delta$ are the heliocentric and geocentric distances expressed in AU and $\Phi(\alpha) \le 1$ is the phase function measured at phase angle $\alpha$.  Equation (\ref{abs}) defines the  absolute  magnitude,  $H$, as the magnitude the comet would have if it could be observed from $r_H = \Delta = 1$ AU and $\alpha =$ 0\degr.  The relevant phase function for Gault is uncertain.  For simplicity, we assume $2.5\log_{10}(\Phi(\alpha))$ = -0.04$\alpha$.  

We further relate the absolute magnitudes  to the effective scattering cross-sections, $C_e$, by 

\begin{equation}
C_e = \frac{1.5\times 10^6}{p_V} 10^{-0.4 H}
\label{area}
\end{equation}

\noindent where $p_V$ is the  geometric  albedo.  The effective albedo of Gault, consisting of a mixture of light scattered from the nucleus and light scattered from dust, is uncertain.  We here assume a nominal albedo $p_V$ = 0.1, while acknowledging that the true value may be larger or smaller by a factor of several.

The catalog value of the absolute magnitude of Gault is $H$ = 14.4 (JPL Horizons software).  This is close to the absolute magnitude in NOT data from February 21  but 0.7 magnitudes fainter than $H$ determined on January 10 (Table \ref{circles}).  Through Equation (\ref{area}), this difference corresponds to $\sim$20 km$^2$ excess in the near-nucleus cross section on January 10.  This estimate is uncertain, however, by an amount that depends on the unknown phase function (and also on the unknown accuracy of the catalog absolute magnitude).  

The apparent magnitudes and effective cross-sections of dust tails A and B are best measured in the high signal-to-noise data from March 03 (Table \ref{geometry}).  We again rotated the composite image to bring each tail to the horizontal, then measured the signal within projected rectangular boxes aligned with each tail.  On Tail A we used a box extending 70\arcsec~along the tail and 10\arcsec~perpendicular, while for the shorter Tail B, these dimensions were 50\arcsec~and 10\arcsec, respectively.  The sky background was determined from the clipped median signal in a series of adjacent regions.  Since the visible tails are longer than the measurable tails, it is clear that the results provide lower limits to the true brightness and cross-section of the ejected dust.  The integrated brightness of Tail is equivalent to $V_A$ = 15.2$\pm$0.2 and of Tail B, $V_B$ = 17.2$\pm$0.3, where the errors reflect the difficulty of measuring low surface brightness diffuse structures against structured backgrounds.  These correspond, through Equation (\ref{abs}), to $H_A$ = 12.3$\pm$0.2 and $H_B$ = 14.3$\pm$0.3 (neglecting additional uncertainties due to the unmeasured phase function). The effective cross-sections (Equation \ref{area}) are $C_A$ = 175$\pm$35 km$^2$ and $C_B$ = 28$\pm$9 km$^2$.  Only very crude photometry of Tail C is possible, because it has extremely low surface brightness.  We estimate $V_C$ = 19.7$\pm$0.7, $H_C$ = 16.8$\pm$0.7 and $C_C$ = 3$\pm$3 km$^2$ from the image on UT 2019 April 10.  For comparison, the geometric cross-section of a 3 km radius sphere is 28 km$^2$.

If contained entirely in spherical  particles of density, $\rho$, and radius, $\overline{a}$, the dust mass is given by $M_d \sim \rho \overline{a} C_e$.  For example, with $\rho$ =  1000 kg m$^{-3}$, $\overline{a}$ = 0.2 mm, the implied dust masses are $M_A \sim 4\times 10^7$ kg, $M_B \sim 6\times 10^6$ kg, and $M_B \sim 6\times 10^5$ kg, respectively.  Taking the event lifetimes for all three tails  as 10 to 20 days, we infer approximate mass production rates $dM_A/dt \sim$ 20 to 40 kg s$^{-1}$, $dM_B/dt \sim$ 4 to 6 kg s$^{-1}$ and $dM_C/dt \sim$ 0.4 to 0.6 kg s$^{-1}$.

\subsection{Colors}
We obtained 3 pairs of BV measurements on UT 2019 January 10 and 2 pairs on UT 2019 February 21, in order to determine the optical colors of Gault.  Measurements of B-V and V-R using circular apertures centered on the photo-center are summarized in Table (\ref{circles}).  Essentially identical central colors are obtained on the two dates, with mean values B-V = 0.77$\pm$0.02, V-R = 0.41$\pm$0.02, redder than the Sun by 0.12$\pm$0.02 and 0.06$\pm$0.02 magnitudes, respectively.  

As noted above, the cross-section in the data from UT 2019 January 10 is dominated by dust, while by February 21 the cross-section is within $\sim$20\% of the value obtained from the JPL Horizons database (which presumably represents Gault in an inactive state).   The fact that the colors remain unchanged even as the scattering cross-section changes from dust dominated to nucleus dominated is evidence that the colors of the nucleus and the dust are similar.  To directly measure the color of the dust free from contamination by the nucleus, we rotated the composite image from UT 2019 January 10 (Figure \ref{image_jan10}) to bring the tail to the horizontal, then measured the signal within projected rectangular apertures, with results  summarized in Table (\ref{boxes}).  In the central or nucleus region, we obtained  colors consistent with those from circular aperture photometry (Table \ref{circles}).   The colors of the dust in Tail boxes A1 and A2 are consistent with those in the central region within the uncertainties of measurement (Table \ref{boxes}), showing that the data provide no evidence for spatial variation of color along the tail of Gault.  Tables (\ref{circles}) and (\ref{boxes}) show that the color of Gault is stable with respect to time and position.

The colors of Gault are compared with those of other asteroid spectral types from Dandy et al.~(2003) in Figure (\ref{bvr}).  Gault is optically too blue to be an  S-type, and is closer to the less metamorphosed members of the C-type complex.  While this makes Gault slightly unusual relative to other members of the Phocaea family, 75\% of which are S-types,  Carvano et al.~(2001) report that 15\% of Phocaea family members are C-types.  Separately, Novakovi{\'c} (2017) identified a low albedo family within the Phocaea group that we tentatively associate with the C-types.

\subsection{Spectrum}
\label{spectra}
The optical reflection spectrum of Gault was obtained on UT 2019 January 20, again at the NOT.  We used a 1\arcsec~wide slit with Grism \#3, which is ruled at 400 line mm$^{-1}$ and gives a dispersion 2.3\AA~pixel$^{-1}$. The resulting spectral resolution $\lambda/\Delta \lambda \sim$ 350.  Two integrations of 1200 s each were obtained, together with a 5 s integration on the V = 7.8 magnitude, G2V solar analog star HIP 53172.  The asteroid and the star were observed at comparable  airmasses of 1.46 to 1.54 (Gault) and 1.52 (star), respectively, in order to cancel both telluric and solar spectral features.  The atmospheric seeing was 0.8\arcsec~to 0.9\arcsec, FWHM.

We extracted the spectrum using a 4.3\arcsec~wide region centered on the peak brightness of the continuum.   After flattening the data using a near-simultaneous spectrum of a quartz lamp, and wavelength calibration using emission lines from an internal HeNe source, the comet and solar analog spectra were divided to produce a reflectivity spectrum. Finally, we removed the slope of the spectrum using a least-squares fit, and normalized the result to unity at $\lambda$ = 5500\AA~(Figure \ref{spectrum}).  We calibrated the continuum using the measured V band (central wavelength $\lambda_c$ = 5500\AA) magnitude, V = 17.82, and the fact that a V = 0 star has flux density $f_{\lambda} = 3.75\times10^{-9}$ erg cm$^{-2}$ s$^{-1}$ \AA$^{-1}$ (Drilling and Landolt 2000).   We estimate that the mean V-band continuum flux density of Gault  is $f_V = 2.8\times10^{-16}$ erg cm$^{-2}$ s$^{-1}$ \AA$^{-1}$.   

Active comets exhibit several emission bands due to resonance fluorescence in cometary gas.  The strongest bands in the optical are from OH (3080\AA)~and CN (3883\AA), neither of which fall within the sensitive range of the NOT spectrograph,  The  strongest band in the 4000\AA~to 7000\AA~range  is from C$_2~\Delta V$ = 0, largely confined to  the range of wavelengths 5050\AA~$\le \lambda \le$ 5220\AA~(c.f.~Farnham et al.~2000).  The  spectrum of Gault shows no evidence for   emission bands at any wavelength.   We define adjacent continuum windows of equal width both blueward (4880\AA~$\le \lambda \le$ 5049\AA, called ``B$_C$'') and redward (5221\AA~$\le \lambda \le$ 5390\AA, ``R$_C$'') of the C$_2$ band in order to assess the noise on the data.  These wavelength bands are indicated in Figure (\ref{spectrum}).  Expressed as a fraction of the continuum, the statistical uncertainties are found to be $\pm$0.013, $\pm$0.010 and $\pm$0.013 in the B$_C$, C$_2~\Delta V$ and R$_C$ bands, respectively.  We take the largest of these values, namely 1$\sigma$ = 0.013, as the best estimate of the fractional uncertainty on the continuum in the C$_2~\Delta V$ band.  Then, 3$\sigma$ corresponds to a flux density $f_{C2} = 0.039   f_V$ = 6.5$\times10^{-18}$ erg cm$^{-2}$ s$^{-1}$ \AA$^{-1}$ (where we have included a correction of 0.6 for slit losses) and this is our empirical upper limit to the gas flux density.  Over the $\Delta \lambda$ = 170\AA~width of the C$_2~\Delta V$ band, the 3$\sigma$ limit to the flux is $F = f_{C2}\Delta \lambda \le$  1.1$\times10^{-15}$ [erg cm$^{-2}$ s$^{-1}$].  

Provided the coma is optically thin, the number of C$_2$ molecules projected within the slit, $N$, is directly proportional to  $F$, following 

\begin{equation}
N = \frac{4 \pi \Delta^2 F}{g(r_H)}.
\label{number}
\end{equation}

\noindent Here, $g(r_H)$ is the fluorescence efficiency factor of the C$_2~\Delta V$ = 0 band at heliocentric distance $r_H$ AU.  We take $g(1)$ = 2.2$\times10^{-13}$ erg s$^{-1}$ radical$^{-1}$ (A'Hearn 1982) and scale to the heliocentric distance of Gault by the inverse square law, finding $g(2.446)$ = 3.7$\times10^{-14}$ erg s$^{-1}$ radical$^{-1}$.  Substituting into Equation (\ref{number}) gives $N \le$ 4$\times10^{26}$ as the 3$\sigma$ upper limit to the number of C$_2$ radicals in the projected 1.0\arcsec$\times$4.3\arcsec~slit. 

To scale from $N$ to the production rate of $C_2$ we need a model to estimate the fraction of the $C_2$ coma captured within the projected slit.  Following cometary tradition, and while recognizing the flaws and uncertainties of the procedure, we use the Haser model formalism, in which the distribution of the radical is assumed to be described by two exponentials with scale lengths characterizing the ``parent'' and the ``daughter'' species.  We use scale lengths at $r_H$ = 1 AU of $L_P = 2.5\times10^4$ km and $L_D = 1.2\times10^5$ km (Cochran 1985), and scale to the heliocentric distance of Gault using $L_{P,D}(r_H) = L_{P,D}(1) r_H^2$.  The speed of outflowing gas molecules, needed to estimate the residence time in the projected slit, was taken to be 0.5 km s$^{-1}$.  

Finally, we find that the non-detection of the C$_2~\Delta V$ = 0 band corresponds to a production rate $Q_{C2} \le 1\times10^{23}$ s$^{-1}$.  In comets, the median ratio of C$_2$ to hydroxyl production rates is C$_2$/OH = 10$^{-2.5}$, albeit with a tail extending to values as low as 10$^{-4}$ (A'Hearn et al.~1995).  The median value would give a spectroscopic limit on the OH production rate limit $Q_{OH} \le 3\times10^{25}$ s$^{-1}$ and mass loss rates in water of $dM/dt \le$ 1 kg s$^{-1}$.  This limit is comparable to values set spectroscopically on active asteroids (e.g.~Jewitt~2012), smaller than the dust production rates inferred in \textsection{ \ref{phot}},  and 10$^2$ to 10$^3$ times smaller than commonly measured in short period comets.  At $r_H$ = 2.446 AU, a perfectly absorbing water ice surface oriented perpendicular to the Sun-Gault line would sublimate in thermodynamic equilibrium at the specific rate $f_s = 5.6\times 10^{-5}$ kg m$^{-2}$ s$^{-1}$.  The empirical spectroscopic limit then sets a limit to the area of exposed ice $f_s^{-1} dM/dt \le 18000$ m$^2$ (0.02 km$^2$).

\section{DISCUSSION}

\subsection{Dust Properties}
The position angles of the tails (Table \ref{geometry}) are plotted as functions of time in Figure (\ref{PAplot}). Lines in the Figure show the position angles of synchrones computed for a range of dates for each tail.  For Tail A, the synchrones span the period UT 2018 October 05 to November 14, with an interval of 10 days.  For Tail B, the synchrones span UT 2018 December 19 to 2019 January 3, with an interval of 5 days.  Evidently, the position angles of the Gault tails are inconsistent with any single isochrone, but compatible with isochrones for dust emitted over discrete ranges of dates.   The middle times for the isochrone models of Tails A and B are UT 2018 October 30 and UT 2018 December 28, respectively, each with an uncertainty of a few days.  Results for Tail C (not plotted in Figure \ref{PAplot}) are less certain, both because the tail is very faint and difficult to measure and because we so far possess only one measurement from UT 2019 April 10.  Our best estimate of the initiation date is UT 2019 February 10$\pm$7.  

\begin{figure}[ht]
\centering
\includegraphics[width=0.95\textwidth]{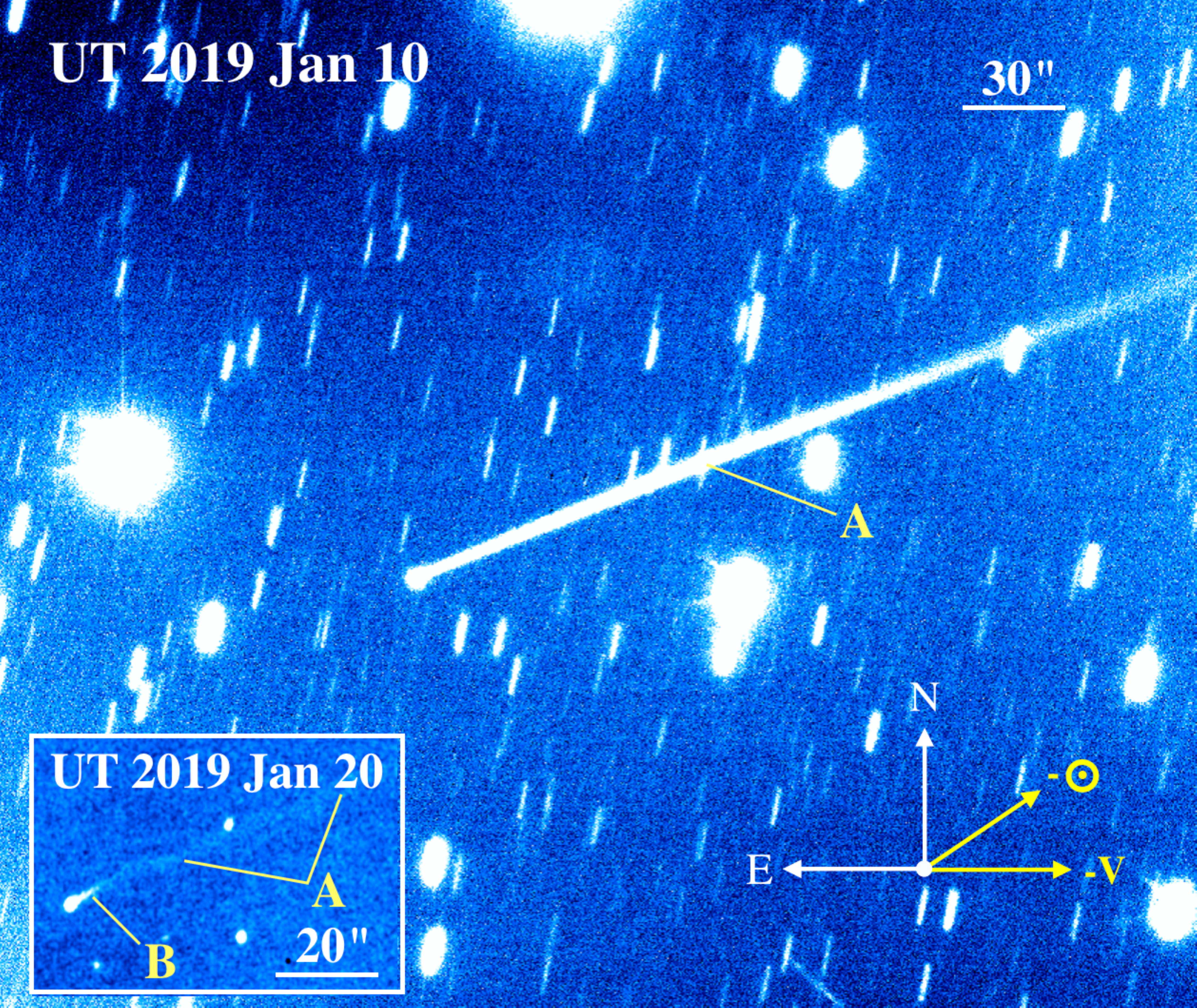}
\caption{Tail A of 6478 Gault on UT 2019 January 10, produced by combining 97 separate 70 s integrations (total exposure time 6790 s) with the NOT.  Cardinal directions, a scale bar and the antisolar and negative velocity vectors are shown. The inset shows Gault on UT January 20, when new Tail B (angled slightly more steeply than the main Tail A) is first visible.  \label{image_jan10} }
\end{figure}

\begin{figure}[ht]
\centering
\includegraphics[width=0.95\textwidth]{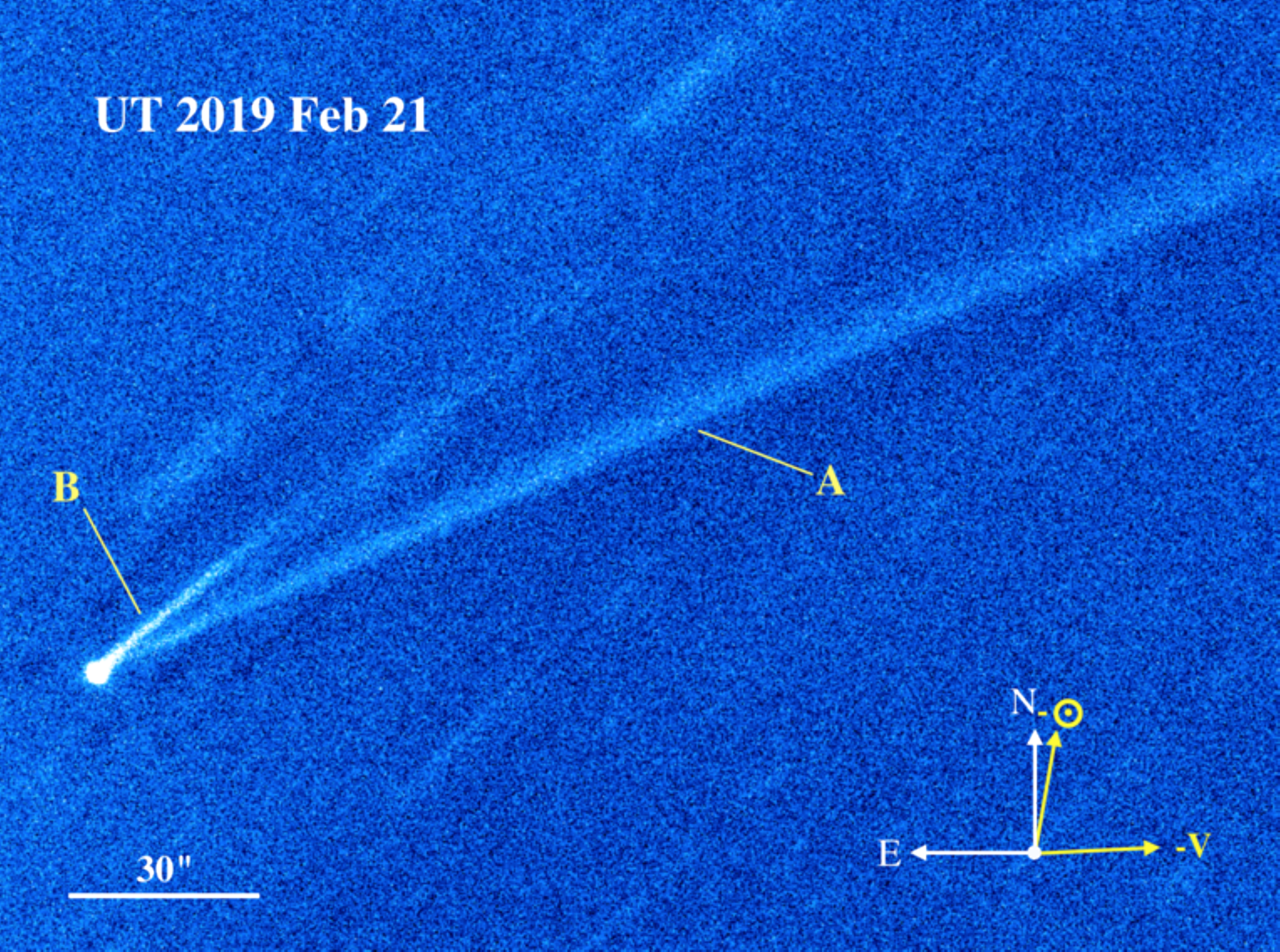}
\caption{6478 Gault on UT 2019 February 21, produced by combining 36 separate 70 s integrations (total exposure time 2520 s) with the NOT. Cardinal directions, a scale bar and the antisolar and negative velocity vectors are shown. Background star and galaxy trail residuals are apparent. \label{image_feb21} }
\end{figure}

\begin{figure}[ht]
\centering
\includegraphics[width=0.95\textwidth]{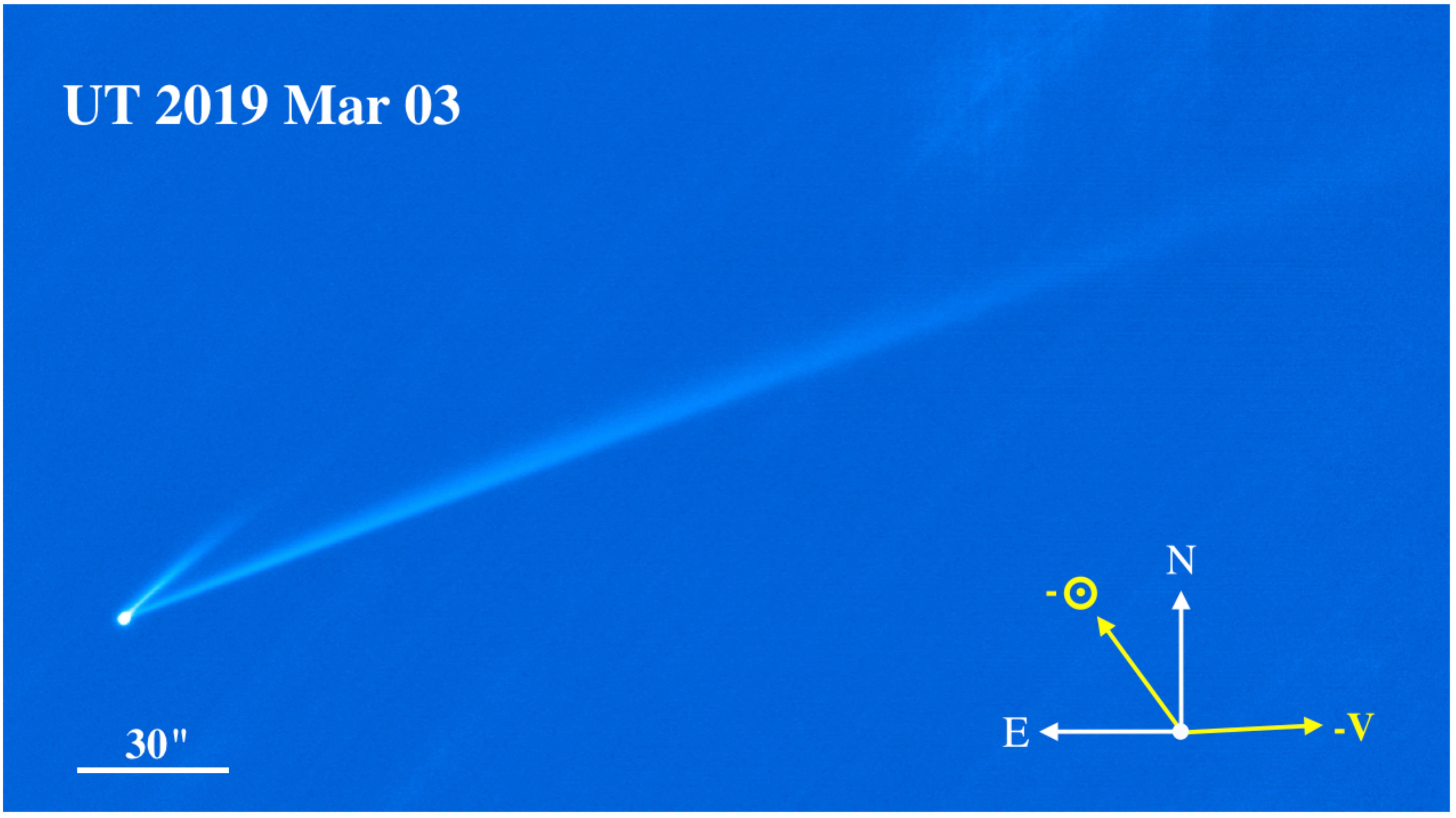}
\caption{6478 Gault on UT 2019 March 03, produced by combining 35 separate 180 s integrations (total exposure time 6300 s) from the WIYN telescope. Cardinal directions, a scale bar and the antisolar and negative velocity vectors are shown. \label{image_mar03} }
\end{figure}

These results imply that Tail A was $\sim$72 days old when imaged on January 10 (Figure \ref{image_jan10}) while Tail B was $\sim$23 days old when first recorded in the image from January 20 (inset to Figure \ref{image_jan10}).  Tail A extends to the edge of the field of view in Figure (\ref{image_jan10}), an angular distance of 250\arcsec.  At geocentric distance $\Delta$ = 1.83 AU (Table \ref{geometry}) this corresponds to a linear distance $\ell = 3.3\times10^8$ m in the plane of the sky.   Particles  released from the nucleus with zero initial velocity and accelerated therafter by radiation pressure would travel a distance  $\ell = \beta g_{\odot}(1) (\Delta t)^2/(2 r_H^2)$, where $r_H$ is the heliocentric distance expressed in AU and $g_{\odot}$(1) = 0.006 m s$^{-2}$ is the gravitational acceleration to the Sun at $r_H$ = 1 AU.  $\Delta t$ is the time since ejection; we set $\Delta t$ = 72 days (6.2$\times 10^6$ s).  Dimensionless quantity $\beta$ is the ratio of the acceleration due to radiation pressure to the acceleration due to solar gravity, and is a function of the particle shape, porosity, density and composition but, most importantly, of size  (Bohren and Huffman 1983).  These parameters are unknown for Gault.  For convenience, we assume that $\beta = 10^{-6}/a$, with $a$ in meters, which is true for grains of density $\rho$ = 570 kg m$^{-3}$.  Substituting, we find $\beta = 0.02$, corresponding to $a \sim 50~\mu m$.  Since the tail is unlikely to lie in the plane of the sky, $\ell$ must be considered a lower limit to the true length, and so $\beta \gtrsim 0.02$ and $a \lesssim 50~\mu m$.   (For example, on January 10, the phase angle was $\alpha = 20\degr$ (Table \ref{geometry}).  If the tail were aligned with the antisolar direction, the actual length would be larger than the plane-of-sky length by a factor $1/\sin(\alpha) \sim 3$, giving $\beta$ three times larger and $a$ three times smaller than our nominal estimate). Larger particles lurk closer to the nucleus.  For example, 25\arcsec~down the tail the particle size suggested by radiation pressure acceleration is $a \sim 500~\mu$m.  The same argument applied to Tail B, which has plane-of-sky length $\sim$ 15\arcsec~on UT 2019 January  20 (Figure \ref{image_jan10}), gives $\ell \gtrsim 2.0\times10^7$ m and $\beta \gtrsim 0.1$, and $a \lesssim 10 \mu$m.  These order of magnitude considerations indicate that the particles ejected from Gault are all very large compared to the wavelength of light.  We estimate a mean particle radius in Tail A $\overline{a} \sim 100 \mu$m.

\begin{figure}[ht]
\centering
\includegraphics[width=0.95\textwidth]{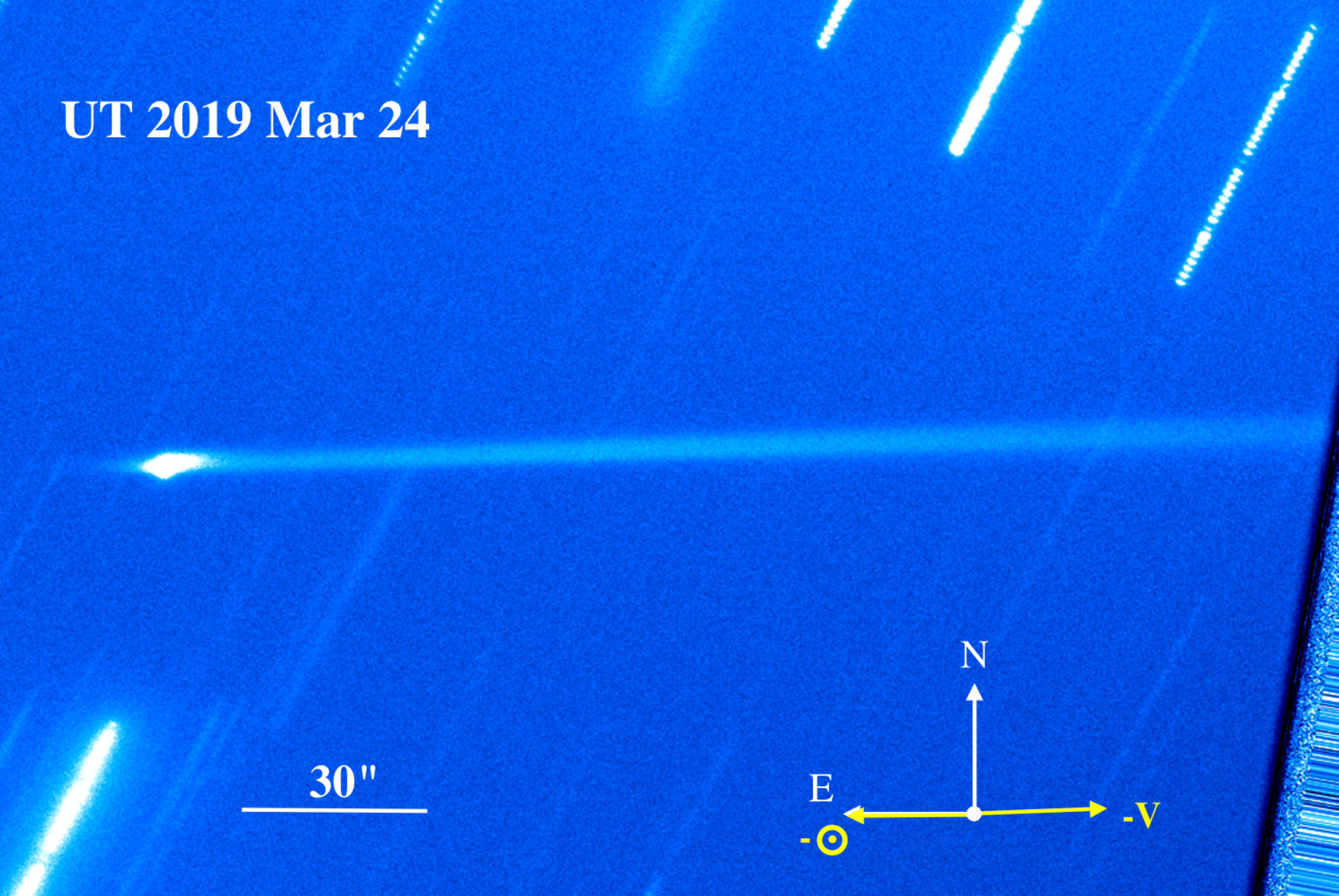}
\caption{6478 Gault on UT 2019 March 24, produced by combining 37 separate 60 s integrations (total exposure time 2220 s) from the NOT telescope. Cardinal directions, a scale bar and the antisolar and negative velocity vectors are shown. \label{image_mar24} }
\end{figure}

\begin{figure}[ht]
\centering
\includegraphics[width=0.95\textwidth]{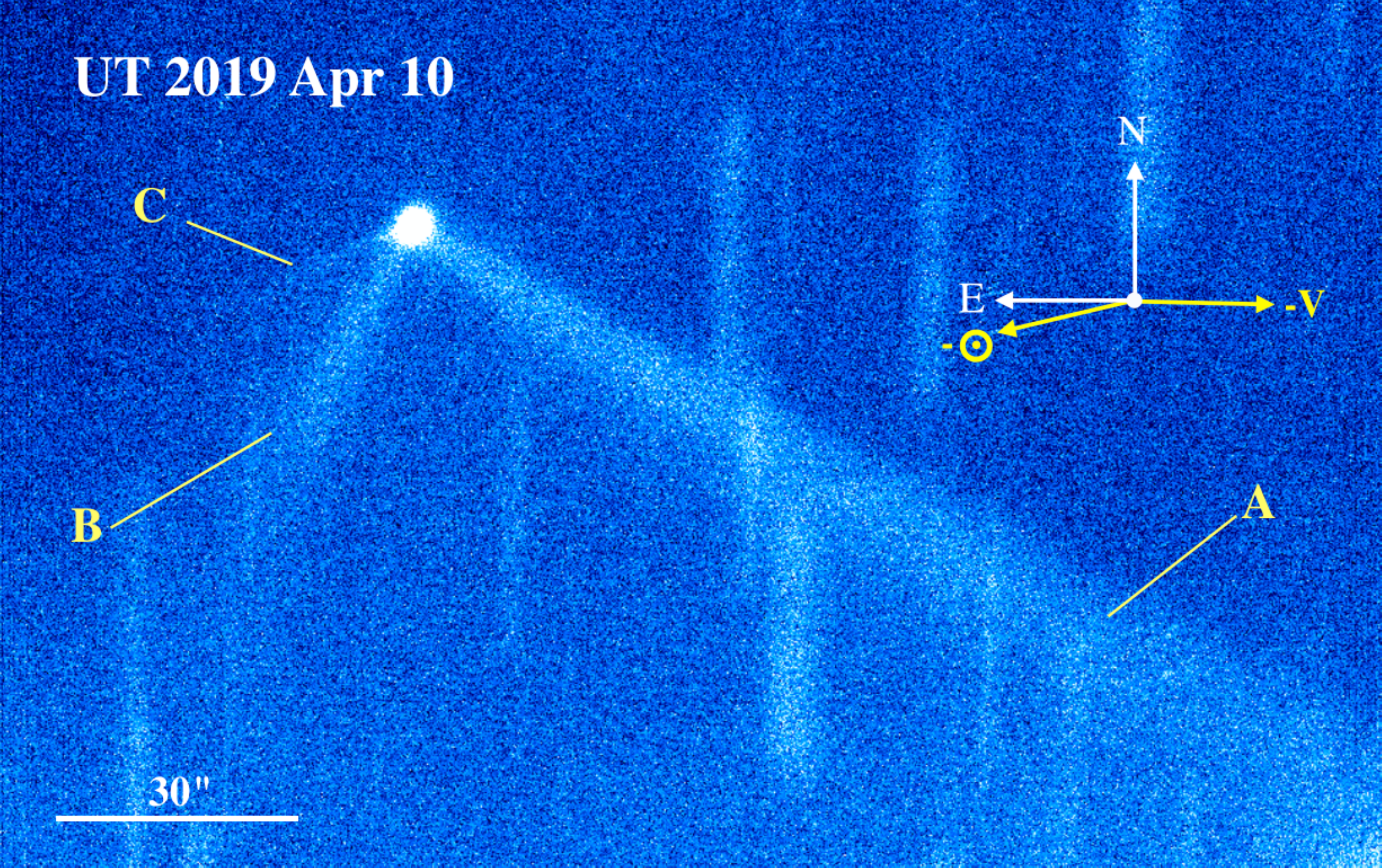}

\caption{6478 Gault on UT 2019 April 10, produced by combining 30 separate 60 s integrations (total exposure time 1800 s) from the NOT telescope. Cardinal directions, a scale bar and the antisolar and negative velocity vectors are shown. \label{image_apr10} }
\end{figure}

\begin{figure}[ht]
\centering
\includegraphics[width=0.95\textwidth]{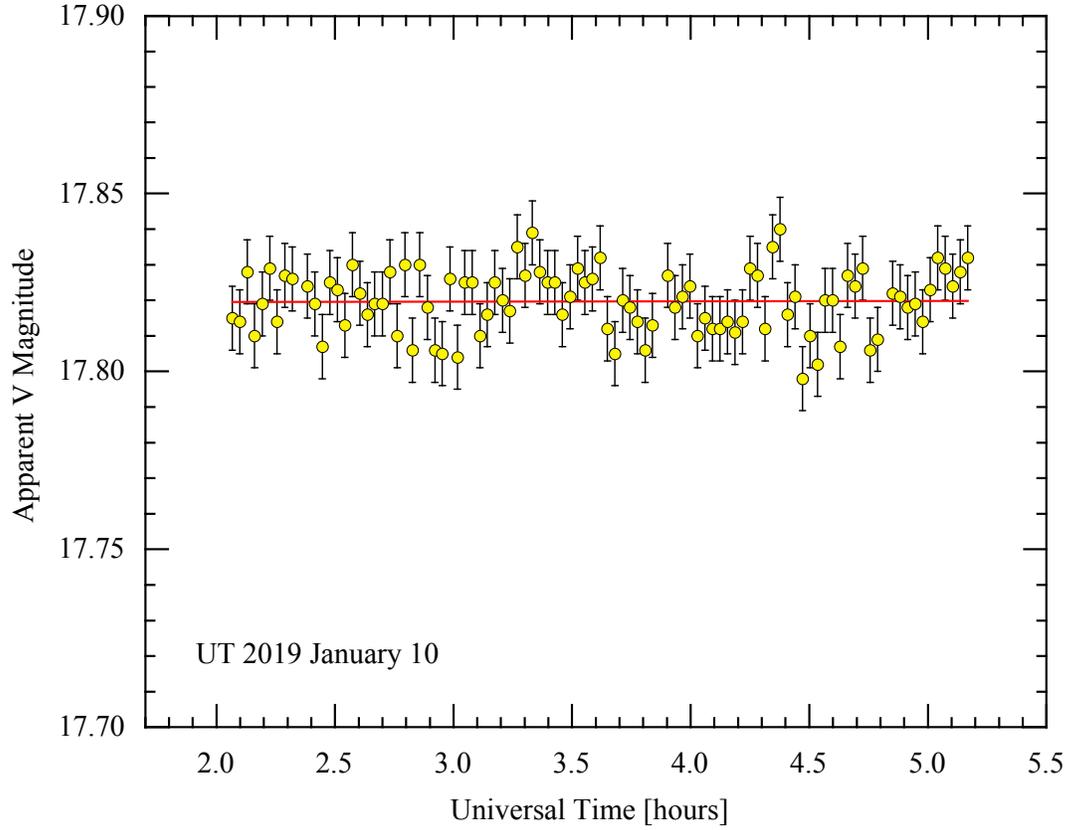}
\caption{Lightcurve measured within a 20 pixel (4.3\arcsec) radius circular aperture with sky subtraction from a surrounding contiguous annulus extending to 8.6\arcsec.  The red line is a linear least-squares fit to the data added to guide the eye.  Representative error bars of $\pm$0.01 magnitudes are shown.  \label{lightcurve} }
\end{figure}

\begin{figure}[ht]
\centering
\includegraphics[width=0.95\textwidth]{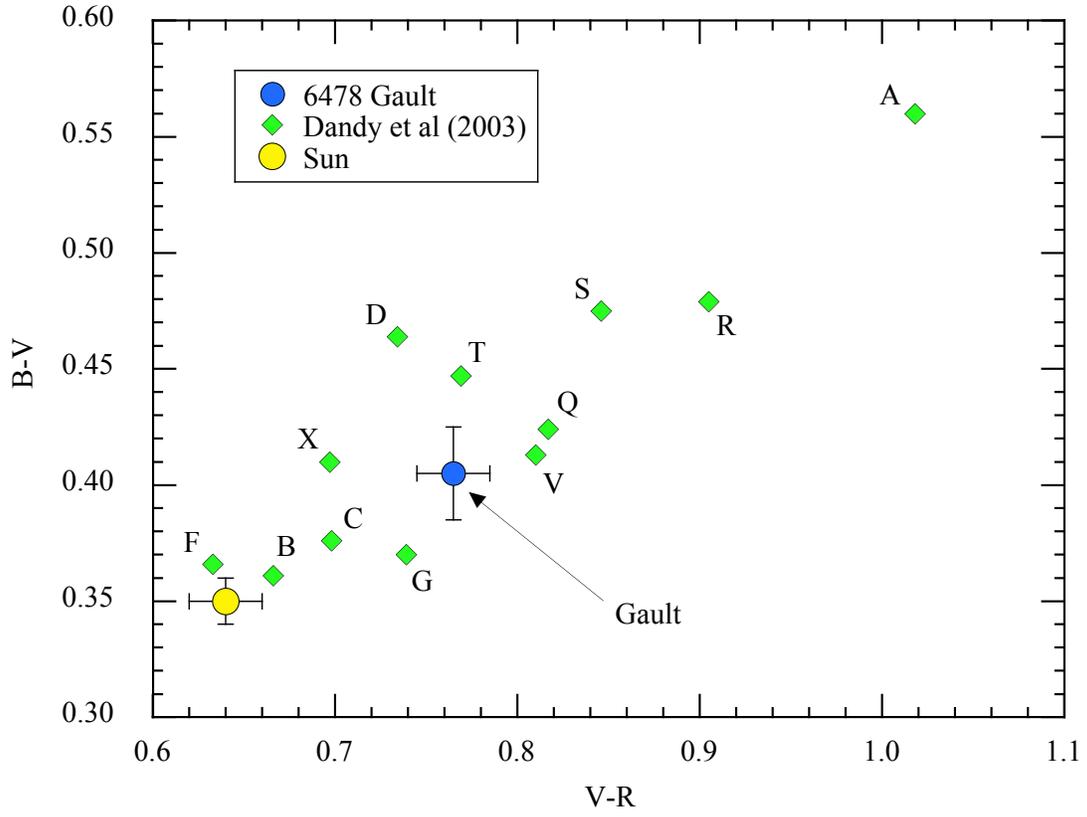}
\caption{Color-color diagram comparing Gault (blue circle with error bars) to various asteroid spectral types from Dandy et al.~(2003), (green diamonds) as marked.  The color of the Sun is indicated by a yellow circle, from  Holmberg et al.~(2006).  \label{bvr} }
\end{figure}

\begin{figure}[ht]
\centering
\includegraphics[width=0.95\textwidth]{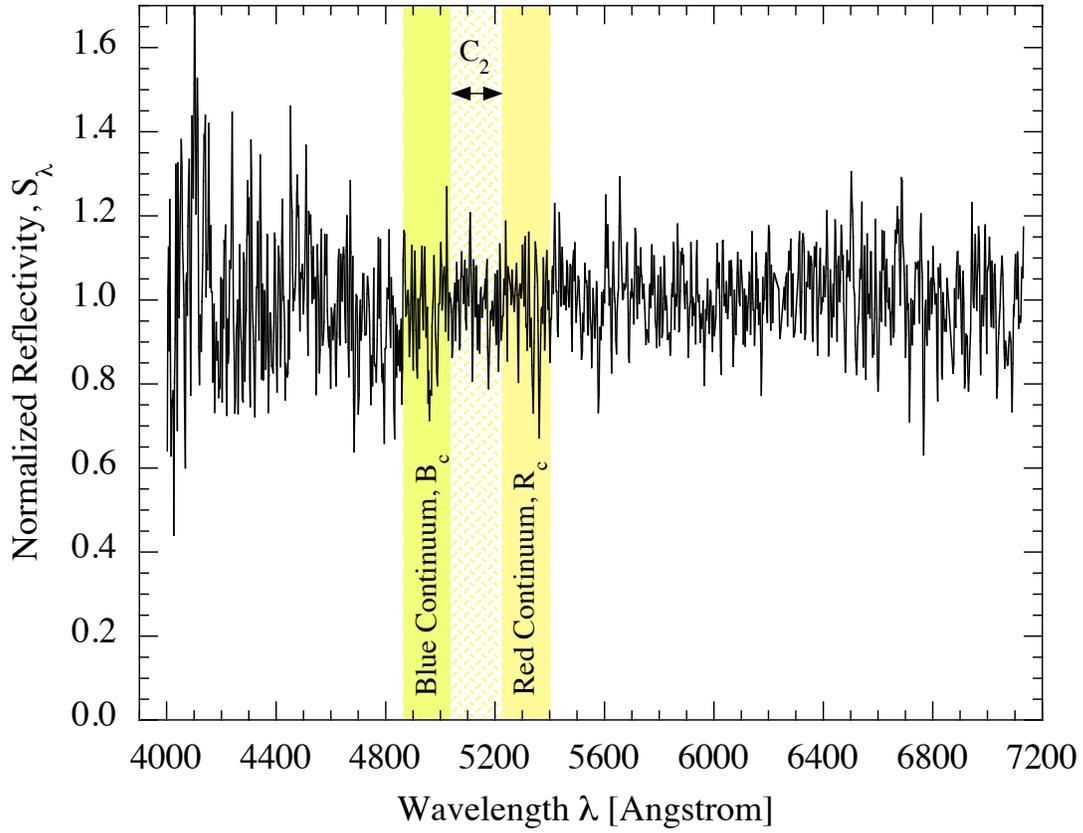}
\caption{De-sloped reflectivity spectrum  normalized to unity at $\lambda$ = 5500\AA, from UT 2019 January 20.  Vertical bands show the wavelengths of the C$_2$ band and flanking blue and red continua (B$_c$ and R$_c$, respectively) used to estimate the noise. \label{spectrum} }
\end{figure}

\begin{figure}[ht]
\centering
\includegraphics[width=0.99\textwidth]{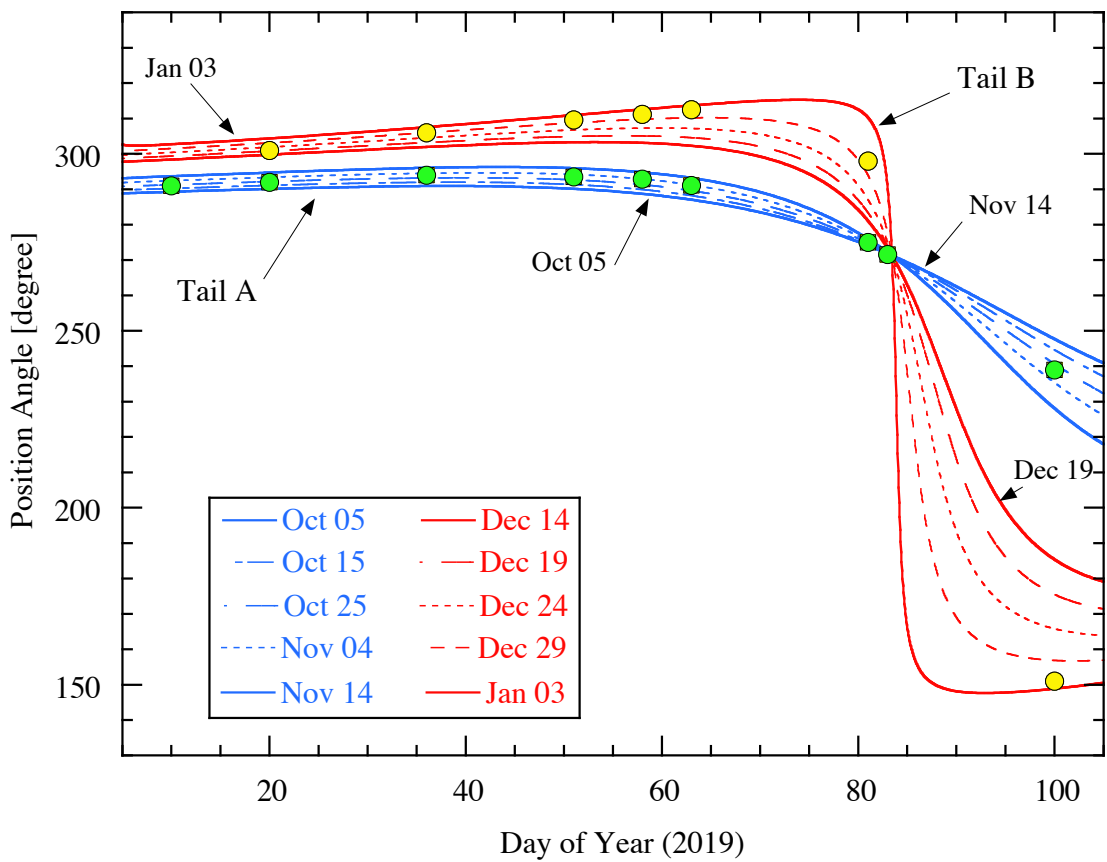}
\caption{Position angles of Tails A and B measured as a function of time, expressed as Day-of-Year (DOY = 1 corresponds to UT 2019 January 1).  The overplotted lines show synchrones for different ejection dates.  We show five synchrones for each tail, with a spacing of 10 days for Tail A in the range October 15 to November 9 and 5 days for Tail B in the range December 19 to January 3. Error bars are smaller than the plot symbols.  \label{PAplot} }
\end{figure}

\begin{figure}[ht]
\centering
\includegraphics[width=0.95\textwidth]{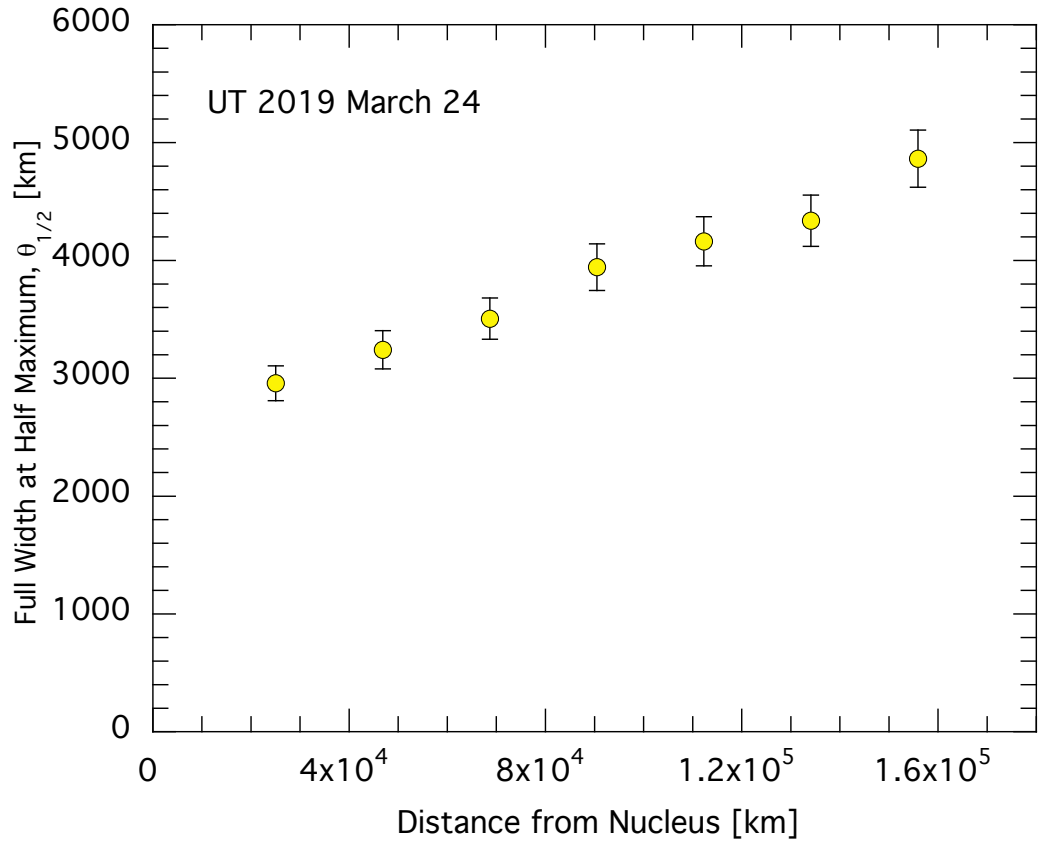}
\caption{Full width at half maximum, $\theta_{1/2}$, of the tail measured on UT 2019 March 24 from a position 0.26\degr~above the orbital plane of Gault.   \label{width} }
\end{figure}

Figure (\ref{image_mar24}) shows Gault on UT 2019 March 24, from a perspective only $\delta_E$ = 0.26\degr~above the orbital plane.  With this viewing geometry, the effects of projection are minimized, although not eliminated, and we obtain a measure of the extent of the dust tail perpendicular to the plane of the orbit.  Notice that Tails A and B appear merged, indicating that they both lie close to the orbital plane of Gault.   To measure the FWHM, $\theta_{1/2}$, we rotated the image to bring the dust tail to the horizontal and then made a series of vertical cuts across the tail, each 100 pixel (21.5\arcsec) wide.  The results are shown in Figure (\ref{width}).  The width increases slowly with distance from the nucleus.  At its narrowest, $\sim$25,000 km from the nucleus, we find $\theta_{1/2}$ = 2.9\arcsec$\pm$0.3\arcsec, corresponding to  3000$\pm$300 km.  The tail is broad compared to the $\theta$ = 1.0\arcsec~FWHM seeing in the Figure (\ref{image_mar24}) image composite, indicating that it is fully resolved.    
To test this, we also measured the higher resolution HST data taken UT 2019 March 22 from a larger angle ($\delta_E$ = 0.7\degr, Table \ref{geometry})  finding a similar width ($\theta_{1/2}$ = 2.8\arcsec$\pm$0.3\arcsec).  

We  use $\theta_{1/2}$ to make a simple estimate of the dust ejection velocity from Gault.  Given that this location samples primarily Tail A, which was ejected on UT 2018 October 28, we infer that the dust age is $\Delta t$ = 1.2$\times10^7$ s.  The half-width then gives an estimate of the ejection speed perpendicular to the orbit as $V_{\perp} = \theta_{1/2}/(2\Delta t)$.  Substituting, we find $V_{\perp} = 0.13\pm0.01$ m s$^{-1}$.  This is an upper limit to the speed because the measured width in Figure (\ref{width}) is potentially still  affected by projection, even though the out-of-plane angle is only 0.26\degr.   The escape speed from a non-rotating 3 km radius sphere of density $\rho$ = 10$^3$ kg m$^{-3}$ is $V_e \sim 2$ m s$^{-1}$.  Escape with $V_{\perp} < V_e$ is consistent with rotation near break-up.

We separately used a Monte Carlo model (developed by Ishiguro et al.~2007) to estimate the dust properties.   The advantage of the model is that it allows the influence of specific assumed dust parameters to be systematically explored.  The disadvantage is that the number of relevant parameters is very large and the model is under-constrained by the data.  We proceeded to adjust the model iteratively in order to test the sensitivity of the tail morphology to different dust and ejection parameters.  

The motion of ejected particles is affected by their initial velocity, $v_0$, by their direction of ejection, and by the magnitude of radiation pressure factor $\beta$.  The gross morphology of the tail is also influenced by the time profile of the emission.  Inspection of the rising portions of the maxima in Figure 2 of Ye et al.~(2019) shows that dust emission continued over 10 to 20 days for both Tails A and B.  We parametrised the emission for each tail as an exponential, $dM/dt \propto \exp(-(t - t_0)/\tau)$, where $t_0$ is the initiation time and $\tau$ is the e-folding time.   We assumed that the velocity vs.~radius relation is of the form $v(a) = v_0 a^{-\gamma}$, with $\gamma$ constant, and we tested two values; $\gamma$ = 1/2 describes particle acceleration against gravity by gas drag while $\gamma$ = 0 describes size-independent ejection, as might be expected from rotational breakup.  We could not find agreeable solutions using $\gamma$ = 1/2 and, further motivated by the very slight measured flaring of the dust tail when viewed edgewise (Figure \ref{width}), adopted $\gamma$ = 0 for all models.  It is conventional to assume that the particles are distributed in size such that the number of particles with radius in the range $a$ to $a + da$ is $n(a) da = \Gamma a^{-q} da$, where $\Gamma$ and $q$ are constants.  However, we found that acceptable fits to the data could not be obtained using this single power law relation, regardless of the choice of $q$.  Instead, we introduced a double power law to better fit the data.   

\begin{figure}[ht]
\centering
\includegraphics[width=0.99\textwidth]{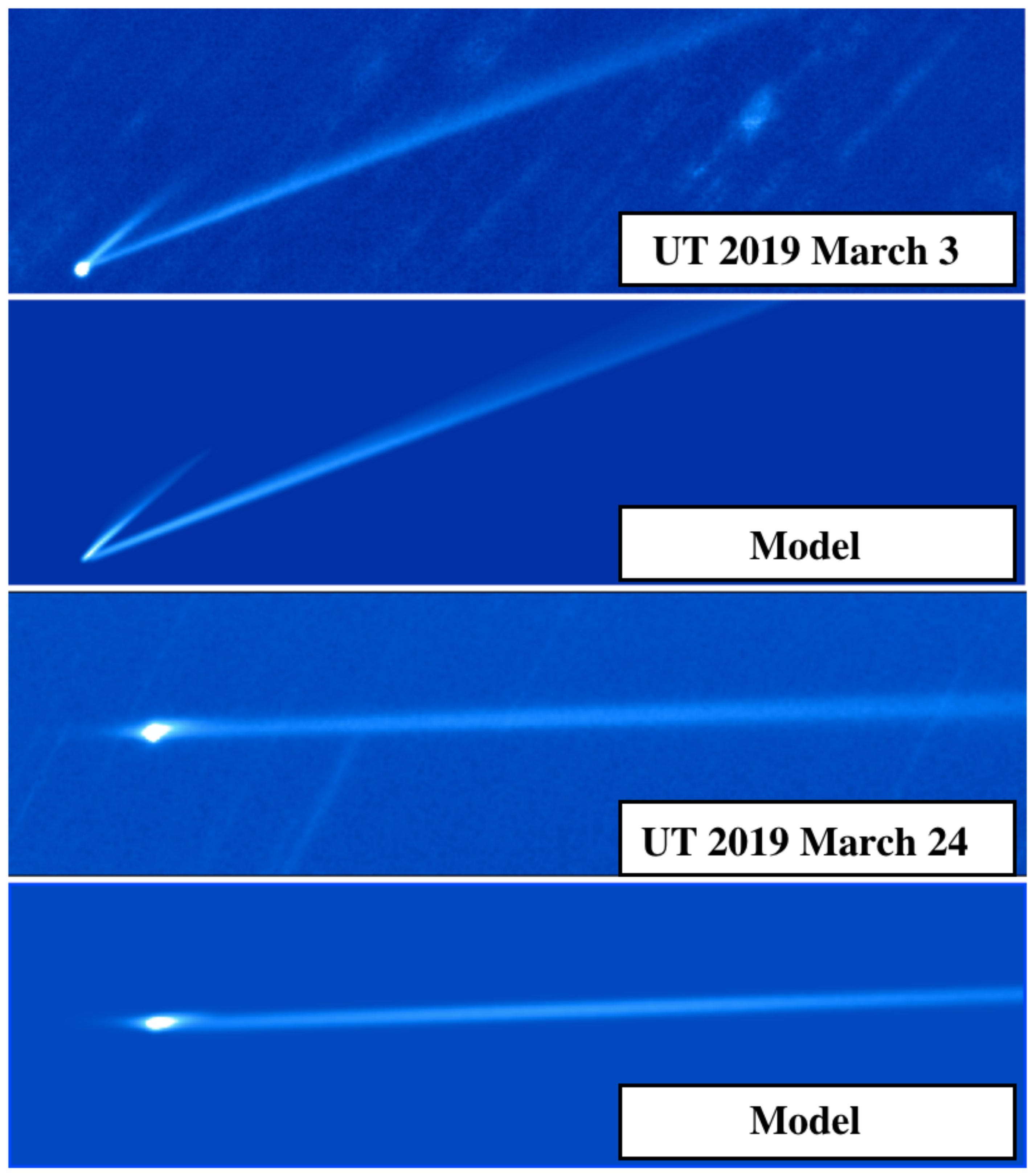}
\caption{Data and simulations of Gault for  (top two panels) UT 2019 March 03 and (bottom two panels) March 24, using the parameters listed in Table (\ref{mcmodels}). \label{models} }
\end{figure}

For both tails A and B, we found acceptable solutions for $v_0$ = 0.15$\pm$0.05 m s$^{-1}$, in excellent agreement with our estimate (0.13$\pm$0.01 m s$^{-1}$) based on analysis of the plane-crossing data.  For Tail A,  ejection started on $t_0$ = UT 2018 October 28 and had an e-folding time $\tau_A$ = 10 days.  For Tail B we found $t_0$ = UT 2018 December 31 and  fading time  $\tau_B \sim$ 5 days.  For both tails, the size distribution in the range 40 $\le a \le 120~\mu$m has $q$ = 3.0$\pm$0.3 while for 120 $\le a \le 5000~\mu$m we find a steeper $q$ = 4.2$\pm$0.2.  The overall similarity of the solutions suggests that the same physical process is responsible for the loss of material in Tails A and B.   The cross-section weighted mean particle radius in this distribution is $\overline{a} = 210$ \micron, with an uncertainty that is at least a factor of two.   We note that these parameters are broadly consistent with the isochrones in Figure (\ref{PAplot}).  Initiation times for Tail A, UT 2018 October 18$\pm$5 and Tail B, UT 2018 December 24$\pm$1 deduced photometrically (Ye et al.~2019) are both slightly earlier than deduced from the dust tail model.   The resulting image-plane models are shown in Figure (\ref{models}). 

The size distribution found by Ye et al.~(2019) is qualitatively similar to that found here, with $q$ = 3.0 for 10 $\le a \le$ 20 \micron\ and $q$ = 4.0 for larger particles. Moreno et al.~(2019) reported three power-law segments with indices $q$ = 2.28 ($1 \le a \le$ 15 \micron), 3.95 (15 $\le a \le$ 870 \micron) and 4.22 ($a >$ 870 \micron) and employed a $v \propto a^{-\gamma}$ relation with $\gamma$ = 1/2, as appropriate for gas drag ejection (although they considered the presence of ice to be unlikely, as do we).  A significantly flatter power-law index, $q \sim 1.6$, was reported by Kleyna et al.~(2019).  However, the latter assumed a single power-law distribution from 10 to 10$^3$ \micron\ and did not specify their choice of $\gamma$.  The models are broadly consistent in their identification of dominantly large particles ejected at speeds comparable to or less than the escape speed from Gault.

\subsection{Origin of the Activity}

\begin{figure}[ht]
\centering
\includegraphics[width=1.00\textwidth]{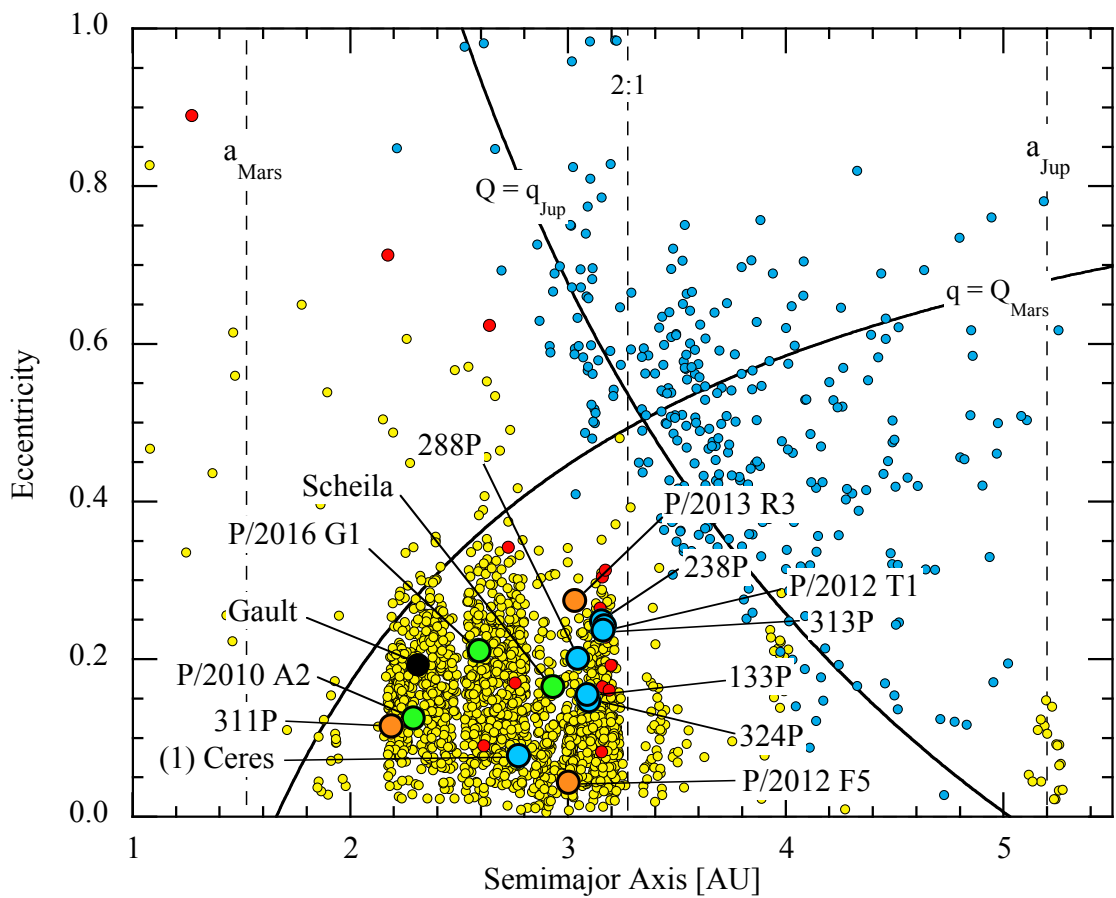}
\caption{Semimajor axis vs.~orbital eccentricity for main-belt asteroids (small yellow circles), comets (small blue circles) and active asteroids.  The latter are color coded by their suspected physical origin; ice sublimation  (filled blue circles), impact debris (filled green circles), rotational breakups (filled orange circles) and other, mostly unknown origins (filled red circles).  Gault is indicated by a filled black circle.  Diagonal arcs show the loci of orbits having perihelion distances equal to the aphelion distance of Mars, and aphelion distances equal to the perihelion distance of Jupiter, as marked.  Vertical dashed lines indicate the semimajor axes of Mars, Jupiter and the 2:1 mean-motion resonance with Jupiter.  Some active asteroids whose symbols overlap at the scale of the plot have been slightly displaced for clarity. \label{ae} }
\end{figure}

The active asteroid closest in  semi-major axis to  Gault is 354P/(2010 A2) (see Figure \ref{ae}), the photometric and morphological properties of which are all indicative of asteroid-asteroid impact (Jewitt et al.~2010, 2013, Kim et al.~2017a, 2017b).  Other examples of likely impact-produced active asteroids are shown in Figure (\ref{ae}) as green-filled circles.  However, we reject impact as a likely cause of mass loss in Gault  for two reasons.  First,  both the position angle  data (Figure \ref{PAplot}) and the photometry (Ye et al.~2019) indicate non-impulsive emission occurring over intervals of $\sim$10 to 20 days, whereas impact should be impulsive.  Second,  the existence of three widely-spaced tails cannot be reconciled with a single impact in any but the most contrived way, and the probability of three consecutive  impacts  is negligible.  

Several examples of ice-sublimating active asteroids are known. However, these differ from Gault by ejecting dust smoothly near perihelion (true anomalies $\nu \sim 0\degr\pm90\degr$, see Hsieh et al.~2018) and do not show the discrete emission required to explain Tails A and B.  Specifically, it is not obvious why exposed ice would sublimate in  fortnight-long windows separated by intervals of months, and Gault is far from perihelion, with true anomaly 240\degr~to 260\degr~(Table \ref{geometry}).  In addition, the isothermal blackbody temperature in Gault's inner-belt orbit is 180 K, reducing (but, admittedly, not eliminating) the prospects that water ice could survive.  Indeed, the best-established ice sublimators are located in the outer belt (filled blue circles in Figure \ref{ae}), where temperatures are lower.  For these reasons we reject ice sublimation as a likely cause of activity in Gault.  We also note that the spectra show no evidence for gas (Figure \ref{spectrum}), although our data are insensitive to emissions from the likely mass-dominant volatile H$_2$O.

\begin{figure}[ht]
\centering
\includegraphics[width=0.75\textwidth]{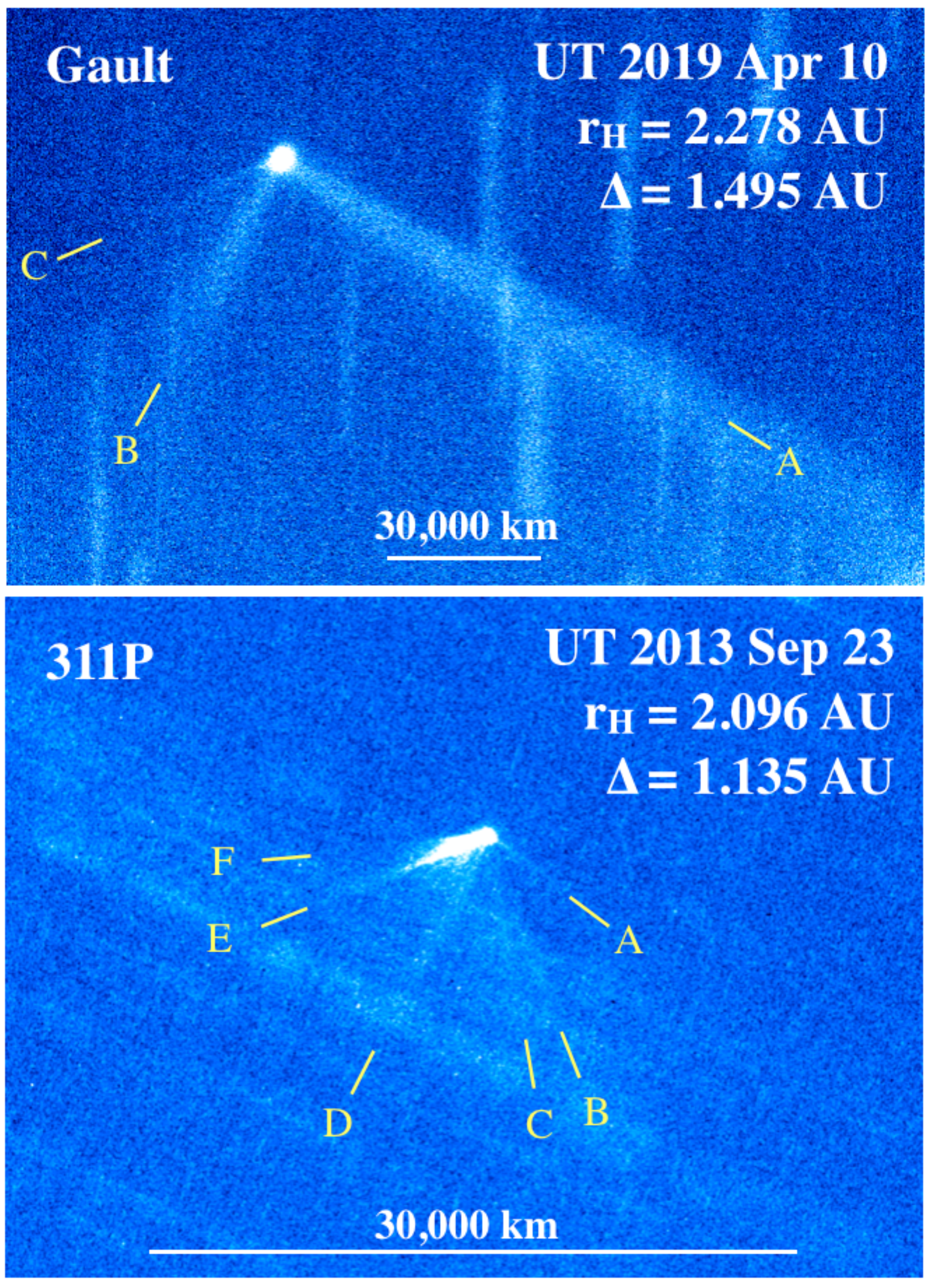}
\caption{Active asteroids  Gault (top panel) and 311P (bottom panel, from Jewitt et al.~2013) compared.  Discrete tails of the two bodies are labelled and a 30,000 km scale bar is shown in each panel.  Gault is physically much larger and brighter than 311P but shares a common appearance, consisting of separate tails coinciding with individual pulses of mass loss, each characterized by a different synchrone.  Both panels have North to the top and East to the left.  \label{starfish} }
\end{figure}

The active asteroid second closest in semi-major axis to Gault in Figure (\ref{ae}) is 311P/(2013 P5).  Figure (\ref{starfish}) shows a comparison between these two objects. Object 311P ejected dust impulsively and aperiodically, producing nine separate tails spread over 9 months (Jewitt et al.~2013, 2015, 2018), with about 10$^5$ kg of dust per ejection.    In between ejections, the nucleus of 311P appears to have been inactive, consistent with the interpretation that each tail formation event signifies the loss of surface fines following local instabilities (``landslides'') on a critically rotating body (Jewitt et al.~2013, Hirabayashi et al.~2015).  Although Gault (so far) has only three tails, the morphological similarity to 311P is striking, leading us to suspect that Tails A, B and C are products of incipient rotational instability on a rapidly rotating nucleus.  One potentially important difference is that 311P is very small, with an estimated radius $\sim$190$\pm$30 m (Jewitt et al.~2018) whereas Gault is $\sim$15 times larger (and $\sim$3500 times more massive), with equivalent circular radius $\sim$2.9 km (from Equation (\ref{area}) with $H$ = 14.4 and $p_V$ = 0.1).  Assuming that neither body contains ice, the torque needed to accelerate the spin presumably comes solar photons and the YORP effect.  The YORP spin-up timescale, all else being equal, is proportional to radius$^2$.  Gault would take 15$^2 \sim$ 225 times longer to reach critical spin than would 311P.  The YORP time for 311P is only $\sim$0.1 Myr (Jewitt et al.~2018), corresponding to 22 Myr when scaled to the size of asteroid Gault.  This longer timescale is still short compared to the  collisional destruction timescale which, for a 3 km radius main-belt body is $\sim10^9$ yr (Bottke et al.~2005).  We conclude that YORP spin-up is capable of driving Gault to incipient rotational instability.   

It is interesting to consider the relative amounts of material shed by 311P and Gault.  A tail mass $M$, if spread uniformly over the parent body of radius $r_n$ would have a thickness, $d$, given by $d = M/(4\pi \rho r_n^2)$.  Substituting, for $M_A$ and $M_B$, we find $d_A = 0.4$ mm and $d_B$ = 0.06 mm.  For comparison, on 311P (with $r_n$ = 0.29 km, Jewitt et al.~2018) the 10$^5$ kg ejected per event corresponds to $d \sim 0.2$ mm, surprisingly similar to the Gault value.  In other words, the ejected mass scales roughly as $r_n^2$, as might be expected for a surface instability.  The reason behind the long duration of mass loss in Gault (e-folding timescales of 5 to 10 days) is less clear.  Our naive expectation is that mass would fall off the surface on the free-fall timescale $(G\rho)^{-1/2} \sim$ hours, rather than days, as observed.  Numerical simulations perhaps provide a hint by showing that, contrary to intuition, unstable systems can survive for days and even years before collisional dissipation and secondary spin fission can lead to the final state  (e.g.~Boldrin et al.~2016).


Sadly, photometry so-far provides no convincing evidence concerning rotation, presumably in part because of the diluting effects of near-nucleus dust.  However, a test for rotational instability should become possible through the acquisition of a rotational  lightcurve once the near-nucleus  dust has cleared.  If the hypothesis is correct, we should find that Gault is rotating near the $\sim$0.1 day critical period for instability (e.g.~Pravec et al.~2008).  It will probably have a very small lightcurve amplitude, as a result of the ``muffin-top'' morphology observed by radar and by direct imaging in several asteroids likely to have been shaped by rotational losses.  The identification of additional discrete tails, either in prediscovery data or in observations yet to be made, or both, would further strengthen the resemblance to the nine-tailed 311P.

\clearpage

\section{SUMMARY}
We present initial observations of active asteroid 6478 Gault.  We argue that Gault is an analogue of the rotationally unstable asteroid  311P/(2013 P5), in which multiple tails result from episodes of equatorward avalanching, followed by pick-up  of the debris by solar radiation pressure.   Specific observations include the following;

\begin{enumerate}

\item 6478 Gault emitted dust with optical colors more C-like than S-like, in three distinct events.  The first ejection began on UT 2018 October 28$\pm$5, the second on UT 2018 December 31$\pm$5 and the third on UT 2019 February 10$\pm$7, with durations 10 to 20 days, and no measurable emission in between.  

\item The dust particles follow a broken power-law size distribution, with index $q$ = 3.0$\pm$0.3 for $40 \le a \le 120$ $\mu$m and $q$ = 4.2$\pm$0.2 for $120 < a \le 5000$ $\mu$m.  The cross-section weighted mean radius is $\overline{a}$ = 0.2 mm.  We infer dust masses $M_A \sim 4\times 10^7$ kg, $M_B \sim 6\times 10^6$ kg and $M_C \sim 6\times 10^5$ kg.  The dust was ejected slowly, with speeds $\sim$0.15 m s$^{-1}$ and we find average mass loss rates $dM_A/dt \sim$ 20 to 40 kg s$^{-1}$, $dM_B/dt \sim$ 4 to 6 kg s$^{-1}$ and $dM_C/dt \sim$ 0.4 to 0.6 kg s$^{-1}$.

\item We set an upper limit to the production of C$_2$ molecules at $Q_{C2} \le 1\times10^{23}$ s$^{-1}$.  Given a nominal cometary ratio C$_2$/OH = 10$^{-2.5}$, the inferred upper limit to the production rate of water is $\le$ 1 kg s$^{-1}$, corresponding to an upper limit to the area of exposed water ice $<$0.02 km$^2$.

\item Despite the likelihood that rotational instabilities drive the current mass loss, time-series photometry of the nucleus region shows no evidence for rotational or other modulation.

\end{enumerate}

\acknowledgments
    
The data presented here were obtained [in part] with ALFOSC, which is provided by the Instituto de Astrofisica de Andalucia (IAA) under a joint agreement with the University of Copenhagen and NOTSA.  Based in part on observations at Kitt Peak National Observatory, National Optical Astronomy Observatory, which is operated by the Association of Universities for Research in Astronomy (AURA) under a cooperative agreement with the National Science Foundation.    We thank Man-To Hui, Jing Li and the anonymous referee for reading the manuscript, Amanda Anlaug (NOT) and Heidi Schweiker (WIYN) for observing assistance.





\clearpage

\begin{deluxetable}{llcllllclrlrrrr}
\tabletypesize{\scriptsize}
\rotate
\tablecaption{Observing Geometry 
\label{geometry}}
\tablewidth{0pt}
\tablehead{\colhead{UT Date} & \colhead{UT}  & Tel\tablenotemark{a} & Mode\tablenotemark{b}   & \colhead{DOY\tablenotemark{c}} & \colhead{$\nu$\tablenotemark{d}} &  \colhead{$r_H$\tablenotemark{e}} & \colhead{$\Delta$\tablenotemark{f}}  & \colhead{$\alpha$\tablenotemark{g}} & \colhead{$\theta_{- \odot}$\tablenotemark{h}} & \colhead{$\theta_{-V}$\tablenotemark{i}} & \colhead{$\delta_{E}$\tablenotemark{j}} & PA$_A$\tablenotemark{k} & PA$_B$\tablenotemark{k} & PA$_C$\tablenotemark{k} }

\startdata

2019 Jan 10 & 02:04 - 05:12      	& NOT & BVR & 10       & 238.8  & 2.466 & 1.833 & 20.4 & 304.4 & 269.8 & 11.6 & 291.0$\pm$0.5 & -- & ---\\
2019 Jan 20 & 02:11 - 02:53      	& NOT & Sp & 20          &  241.3  & 2.446 & 1.713 & 18.4 & 309.2 & 270.3 & 11.6 &  292.0$\pm$1.0 & 301.0$\pm$1.0 & ---\\
2019 Feb 05 & 15:08 - 15:45 		& HST  & F350LP & 36 &  245.2   &   2.413 & 1.537 & 13.4 & 322.0 & 271.5 & 10.6 & 294.0$\pm$0.5 & 306.0$\pm$0.5 & --- \\
2019 Feb 21 & 00:27 - 01:44 		& NOT & BVR    & 51   &  249.3   &  2.382 & 1.429 & 8.2 & 350.5 & 272.5 & 8.2 & 293.5$\pm$0.5 & 309.6$\pm$0.7 & ---\\
2019 Feb 27 & 16:17 - 16:51		& HST & F350LP & 58 &  250.9   &  2.368 & 1.402 & 6.9 & 15.4 & 272.7 & 6.7 & 292.9$\pm$0.5 & 311.2$\pm$0.5 & ---\\
2019 Mar 03 & 04:43 - 07:17 		& WIYN & r' 		& 63         &  252.2  	& 2.360 & 1.392 & 6.7 	& 31.7 & 272.7 & 5.9 & 291.1$\pm$0.5 & 312.5$\pm$0.5  & ---\\
2019 Mar 22 & 10:53 - 11:29         	& HST & F350LP  	& 81 	& 257.1 	& 2.319 & 1.401 & 12.4 	& 88.4 & 271.5 & 0.70 & 275.0$\pm$0.5 & 298$\pm$2 & ---	\\
2019 Mar 24 & 00:16 - 01:21             & NOT & R                 & 83 & 257.6 & 2.315 & 1.406 & 13.0 & 90.3 & 271.3 & 0.26 & 271.6$\pm$0.2 & --- & --- \\
2019 Apr 09-10 & 23:31 - 00:36 		& NOT & R & 100 & 262.3 & 2.278 & 1.495 & 19.5 &   103.0 & 268.9 & -4.0 & 	239$\pm$1 & 151$\pm$1 & 116$\pm$1  \\

\enddata


\tablenotetext{a}{NOT = NOT 2.5m, HST = Hubble Space Telescope 2.6 m, WIYN = 3.5 m}
\tablenotetext{b}{BVR = imaging at 0.214\arcsec~pixel$^{-1}$, Sp = spectrum, F350LP see text}
\tablenotetext{c}{Day of Year number, 1 = UT 2019 January 01}
\tablenotetext{d}{True anomaly, in degrees}
\tablenotetext{e}{Heliocentric distance, in AU}
\tablenotetext{f}{Geocentric distance, in AU}
\tablenotetext{g}{Phase angle, in degrees}
\tablenotetext{h}{Position angle of projected anti-solar direction, in degrees}
\tablenotetext{i}{Position angle of negative projected orbit vector, in degrees}
\tablenotetext{j}{Angle of Earth above orbital plane, in degrees}
\tablenotetext{k}{Position angles of Tails A, B and C}

\end{deluxetable}

\clearpage

\begin{deluxetable}{lccccccc}

\tablecaption{Circular Aperture Color Photometry 
\label{circles}}
\tablewidth{0pt}

\tablehead{UT Date & Radius\tablenotemark{a} &  V & H\tablenotemark{b}& B-V & V-R & B-R }
\startdata
January 10 & 4.3\ 		& 17.82$\pm$0.02 	& 13.73 & 0.78$\pm$0.03			&   0.40$\pm$0.03	& 1.18$\pm$0.03 \\
February 21 & 4.3\ 		& 17.61$\pm$0.02 	& 14.62  	& 0.75$\pm$0.03		& 0.41$\pm$0.03 	& 1.16$\pm$0.03 \\

\hline
Sun\tablenotemark{c}       &               ---         & ---              & ---   & 	0.64$\pm$0.02 & 0.35$\pm$0.01 & 1.00$\pm$0.02 \\

\enddata

\tablenotetext{a}{Angular radius of circular photometry aperture, in arcseconds.}
\tablenotetext{b}{Absolute magnitude, from Equation (\ref{abs}).  The uncertain is dominated by the unmeasured phase function and is of order several $\times$0.1 magnitudes.}
\tablenotetext{c}{Colors of the Sun  from Holmberg et al.~(2006).}

\end{deluxetable}

\clearpage

\begin{deluxetable}{lrccccc}

\tablecaption{Rectangular Aperture Color Photometry\tablenotemark{a} 
\label{boxes}}
\tablewidth{0pt}

\tablehead{ \colhead{Region\tablenotemark{b}} & $x_{min}, x_{max}$ &  $y_{min}, y_{max}$ & V & B-V & V-R & B-R }
\startdata
Nucleus  			& -2.2, 2.2\ & -2.2, 2.2  & 17.88$\pm$0.01	&   0.78$\pm$0.01	& 0.40$\pm$0.01 & 1.18$\pm$0.01  \\
Tail A    			& 2.2, 44.9  & -2.2, 2.2  & 20.51$\pm$0.02 	&   0.77$\pm$0.08 & 0.42$\pm$0.05 & 1.19$\pm$0.09 \\
Tail A  			& 44.9, 87.9  & -2.2, 2.2 & 20.94$\pm$0.03      &  0.81$\pm$0.10  & 0.52$\pm$0.08 & 1.33$\pm$0.11 \\

\hline
Sun\tablenotemark{c}       &               ---               &          ---          & ---   & 	0.64$\pm$0.02 & 0.35$\pm$0.01 & 1.00$\pm$0.02 \\

\enddata

\tablenotetext{a}{All data from UT 2019 January 10.}
\tablenotetext{b}{Rectangle defined by $x_{min} \le x\arcsec \le x_{max}$,  $y_{min} \le y\arcsec \le y_{max}$, with the nucleus  at $x$\arcsec~= $y$\arcsec~= 0, and  $x$  increasing along the tail to the west }
\tablenotetext{c}{Colors of the Sun  from Holmberg et al.~(2006).}

\end{deluxetable}

\clearpage

\begin{deluxetable}{lcccccc}

\tablecaption{Dust Model Parameters\tablenotemark{a} 
\label{mcmodels}}
\tablewidth{0pt}

\tablehead{ \colhead{Feature} & $v_0$\tablenotemark{b} &   $t_0$\tablenotemark{c}  & $\tau$\tablenotemark{d}  & $q_1\tablenotemark{e}$ & $q_2\tablenotemark{f}$ &   }
\startdata
Tail A    & 0.15$\pm$0.05  	& 2018 October 28$\pm$5 & 10 & 3.0$\pm$0.1	& 4.2$\pm$0.2  	  \\
Tail B    & 0.15$\pm$0.05 	& 2018 December 31$\pm$5 & 5 & 2.7$\pm$0.2 	& 4.2$\pm$0.2 		   \\

\enddata

\tablenotetext{a}{Results from Monte Carlo simulations.}
\tablenotetext{b}{Dust ejection velocity  (m s$^{-1}$)}
\tablenotetext{c}{Tail initiation date}
\tablenotetext{d}{e-folding fading time, in days}

\tablenotetext{e}{Size distribution index for particle radii $40 \le a \le 120~\mu$m}
\tablenotetext{f}{Size distribution index for particle radii $120 \le a \le 5000~\mu$m}

\end{deluxetable}

\clearpage

\clearpage

\clearpage

\clearpage

\clearpage

\clearpage

\clearpage

\clearpage

\clearpage

\clearpage

\clearpage

\clearpage


\end{document}